\documentclass[final,5p,times,twocolumn]{elsarticle}
\newcommand{\pf}[2]{\frac{\partial ,1}{\partial ,2}}
\newcommand{\pfs}[2]{\frac{\partial^2 ,1}{\partial ,2 ^2}}
\newcommand{\pft}[2]{\frac{\partial^3 ,1}{\partial ,2 ^3}}
\newcommand{\pderiv}[2]{\frac{\partial ,1}{\partial ,2}}

\usepackage{multirow}
\usepackage{tabu}
\usepackage{CJKutf8}
\usepackage[T1]{fontenc}
\usepackage[utf8]{inputenc}
\usepackage{graphics}
\usepackage{graphicx,color}
\usepackage{epsfig}
\usepackage{amsmath}
\usepackage{amssymb}

\usepackage{setspace}
\usepackage{hyperref}
\usepackage{cleveref}
\usepackage{underscore}
\usepackage[dvipsnames]{xcolor}
\usepackage{multirow, makecell}
\usepackage{colortbl}
\usepackage{dashrule}
\usepackage{ehhline}
\usepackage{csquotes}
\usepackage{natbib}
\biboptions{sort&compress}

\usepackage{array}
\usepackage{booktabs}
\usepackage{caption}
\makeatletter
\usepackage{colortbl}
\usepackage{hhline}

\journal{International Communication in Heat and Mass Transfer}
\begin{document}
	
	\begin{frontmatter}
		
	\title{Heat Dissipation and Thermoelectric Performance of InSe-Based Monolayers: A Monte Carlo Simulation Study}

\author[1]{Seyedeh Ameneh Bahadori}
\author[1]{Zahra Shomali\corref{cor1}}
	\cortext[cor1]{Corresponding author, Tel.: +98 21 82884785.}
	 \ead{shomali@modares.ac.ir} 
	\address[1]{Department of Physics, Faculty of Basic Sciences, Tarbiat Modares University, Tehran 14115-175, Iran}
\author[2]{Reza Asgari}
\address[2]{School of Qunatum Physics and Matter, Institute for Research in Fundamental Sciences (IPM), Tehran 19395-5531, Iran}

\begin{abstract}
Using nonequilibrium Monte Carlo simulations of the phonon Boltzmann transport equation, we study transient heat transfer in five indium-based two-dimensional monolayers: Janus monolayers In$_2$SeTe and In$_2$SSe, pristine InSe, and InSe under 4$\%$ and 6$\%$ tensile strain. In this work, the potential of these materials for energy conversion in thermoelectric generators and hotspot control in metal-oxide-semiconductor field-effect transistors is investigated. A promising option for an effective heat dissipation and enhanced transistor reliability is found to be a strained InSe, which shows the lowest peak temperature during the heating among the studied materials. On the other hand, with a high Seebeck coefficient, low thermal conductivity, and an improved figure of merit, the Janus In$_2$SeTe monolayer, compensates for its increased phonon scattering to reach the maximum temperature, making it a potent thermoelectric material. Our findings emphasis the importance of strain engineering and structural asymmetry in tuning phonon transport, enabling material optimization for next-generation nanoelectronic and energy-harvesting devices. 
\end{abstract}

		
	\end{frontmatter}

	
\section{Introduction}

 As technological advancements continue to drive the development of more compact, powerful, and highly miniaturized transistor chips, the issue of self-heating within these devices has become increasingly critical and challenging to manage \cite{Sheng2023,AYan2024,HLi2024}. Ineffective heat dissipation not only risks device malfunction and reduced lifespan but also contributes to broader environmental concerns. Consequently, there is a growing demand for efficient thermal management strategies. In this context, thermal management has emerged as a key engineering priority, involving the application of heat transfer principles to maintain the temperature of electronic systems within safe operational limits \cite{Moore2014,Mahajan2002}. In nanoscale electronics, device reliability is often dictated by the temperature at the hottest region of the die, known as the thermal hotspot \cite{Samian2013,Samian2014,Moghadam2014}. By analyzing the thermal behavior of individual transistors or entire chips, these hotspots can be identified and mitigated through targeted solutions such as optimally placing heat spreaders or using materials with superior thermal properties \cite{Subrina2009,Shomali2012,Shomali20152,Shomali2016,Shomali2019,Sattler2022}.
 
Another effective thermal management strategy involves the use of materials that experience minimal temperature rise under self-heating conditions. Developing such thermally efficient materials is critical for advancing MOSFET technology and ensuring device reliability at the nanoscale \cite{Shomali2018,Shomali2023,Bahadori2024}. In parallel, thermoelectric technology represents a promising and complementary approach to thermal management. Thermoelectric energy harvesting offers a dual benefit: it dissipates excess heat while partially recovering waste heat as usable electrical energy. This process relies on the Seebeck effect, where an electric voltage is generated in response to a temperature gradient. When integrated into electronic systems, thermoelectric materials not only improve energy efficiency but also contribute to temperature regulation, offering a sustainable path toward enhanced thermal control.
 
Recent advances in nanotechnology have opened new avenues for thermal management and thermoelectric energy harvesting. In particular, two-dimensional (2D) nanomaterials and nanostructures are being actively explored for their tunable thermal conductivity, making them promising candidates for a wide range of applications \cite{Wan2019,Wu2024}. In our efforts to identify thermally optimal materials for use in thermoelectric devices and MOSFETs, we focus on newly engineered indium-based monolayer compounds with advantageous physical properties. Indium selenide (InSe) monolayers, known for their high electron mobility, tunable bandgap, direction-dependent electronic behavior, and excellent thermal stability, have attracted considerable attention. Moreover, 2D InSe exhibits good ambient stability in nano-device environments \cite{Nan2018}.

InSe monolayers subjected to tensile strain display modified thermal and electrical properties, making them particularly interesting for thermal management. While tensile strain tends to reduce the bandgap and Seebeck coefficient, it also induces stronger anharmonic phonon scattering, lower phonon group velocities, and reduced heat capacity—factors that collectively lead to lower lattice thermal conductivity and thus greater thermoelectric potential.

In addition to strained InSe, Janus monolayers such as In$_2$SSe and In$_2$SeTe have recently emerged as promising candidates in thermoelectric engineering, owing to their dynamic and thermal stability \cite{Vu2021}. These materials possess moderate bandgaps that can be further tuned via strain engineering. For example, p-type doping of the Janus In$_2$SeTe monolayer leads to a substantial enhancement in electrical conductivity. Specifically, its electrical conductivity ($\sigma/\tau$) reaches 10.81$\times$10$^{19}$($\Omega^{-1}$m$^{-1}$s$^{-1}$), representing a significant improvement over its undoped state \cite{Zhang2018,Khengar2022}. This combination of enhanced electrical transport and favorable thermoelectric characteristics establishes In$_2$SeTe as a strong candidate for energy-harvesting applications.

In this work, we perform detailed phonon analysis to identify dominant heat carriers and assess spatial and temporal heat distribution in In-based monolayers. By tracing temperature evolution and heat flux, we evaluate the suitability of each material as either a thermoelectric element or a MOSFET channel. We focus on five monolayers: pristine InSe, InSe under 4$\%$ and 6$\%$ tensile strain, and the Janus compounds In$_2$SeTe and In$_2$SSe. Materials with the lowest peak temperatures are considered most promising for replacing silicon in MOSFETs, due to improved thermal reliability. Conversely, materials that exhibit higher peak temperatures (and thus greater temperature gradients) are better suited for thermoelectric energy harvesting \cite{Bahadori2024}. In summary, we analyze the peak thermal behavior of these five In-based monolayers to identify optimal candidates for thermal management and thermoelectric applications. 

The structure of the paper is as follows: Section \ref{Sec.2} presents the device geometry and boundary conditions; Section \ref{Sec.3} introduces the mathematical modeling, and Section \ref{Sec.4} outlines the numerical method. Results and discussions are provided in Section \ref{Sec.5}, followed by conclusions in Section \ref{Sec.6}.

	\section{Geometry}      
	\label{Sec.2}
Indium selenide (InSe) monolayers consist of four atomic planes arranged in a hexagonal configuration. These layers are held together by covalent bonds in a Se-In-In-Se sequence, where each indium atom is tetrahedrally coordinated with three neighboring selenium atoms. This bonding arrangement results in a layered structure that forms a honeycomb lattice of indium and selenium atoms. InSe exists in three primary polytypes: $\beta$, $\epsilon$, and $\gamma$. Both the $\beta$ and $\epsilon$ phases exhibit hexagonal symmetry and are composed of two-layer unit cells \cite{Politano}. In particular, the structural diversity of $\beta$-InSe enables it to exfoliate into stable, atomically smooth two-dimensional sheets under vacuum conditions. At room temperature, InSe typically crystallizes in the rhombohedral $\gamma$-phase, which belongs to the R3m space group \cite{Dmitriev2011}. The individual layers in these structures are bonded by weak van der Waals interactions \cite{Politano}, allowing for facile exfoliation into nanosheets or nanoparticles. These characteristics position InSe as a promising material for applications in electronics and optoelectronics.

As a semiconductor, InSe possesses a direct bandgap of approximately 1.3 eV and exhibits high electron mobility, reaching values up to 2000 cm$^2$V$^{-1}$s$^{-1}$, surpassing that of silicon and many other transition metal dichalcogenides. Viewed from above, the monolayer adopts a honeycomb lattice with a lattice constant of a$_0$=3.95$\AA$. In this study, mechanical strain is applied to the InSe monolayer to preserve crystal symmetry while enabling the modulation of its thermal properties. The applied tensile strain is quantified by the relation $\gamma$=(a-a$_0$)/a$_0$$\times$100$\%$, where $a$ and a$_0$ denote the strained and unstrained lattice constants, respectively \cite{Wan2020}.

In addition to strained InSe, we examine two Janus derivatives: In$_2$SeTe and In$_2$SSe monolayers. The Janus In$_2$SeTe structure consists of a Te-In-Se sequence in a honeycomb configuration, created by substituting one Se atom in InSe with a Te atom \cite{Ibrarra-Hernández2022,Wan2019}. Similarly, the In$_2$SSe monolayer features asymmetrically bonded S and Se atoms on opposing surfaces, with two indium layers positioned in between.

\begin{figure}
	\vspace{0cm}
	\centering
	\includegraphics[width=0.9\columnwidth]{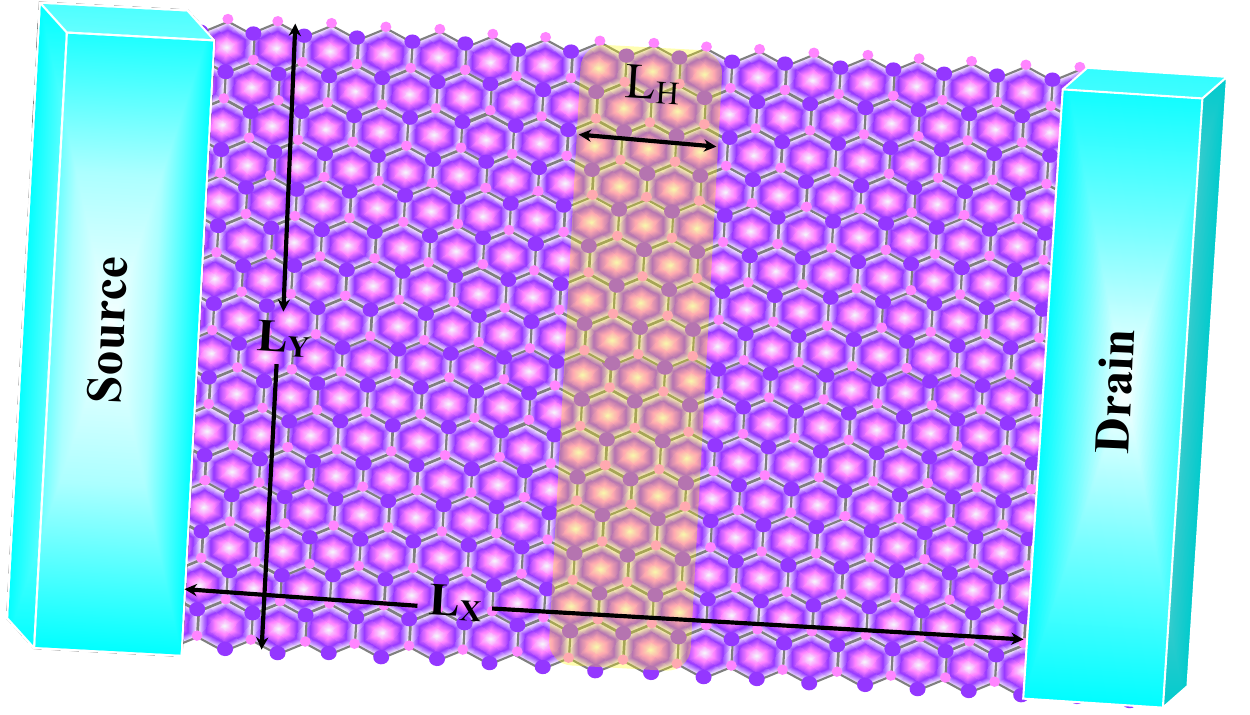}
	\caption{\label{Geometry}The schematic of the MOSFET with channels of InSe, InSe under 4$\%$ and 6$\%$ strain, In$_2$SeTe or In$_2$SSe with L$_x$ and L$_y$ being the width and the length. The heat generation area is indicated by L$_H$.}
\end{figure}

The geometric parameters of the simulated monolayers are as follows: width L$_x$=100 nm, length L$_y$=100 nm, and thicknesses of 0.757 nm (InSe), 0.546 nm (In$_2$SeTe), and 0.539 nm (In$_2$SSe). As illustrated in Fig.\ref{Geometry}, a uniform heat flux of Q=10$^{12}$ W/m$^3$ is applied to the central region of each monolayer channel, specifically between x=$45$ nm and x=$55$ nm along the x-direction. All boundaries are assumed adiabatic, except for the open bottom boundary, which is exposed to the ambient environment. The initial temperature of each structure is set to 299 K.

	\section{Mathematical Modeling}
	\label{Sec.3}

The nonequilibrium Monte Carlo (NMC) method is used to simulate the phonon Boltzmann transport equation (PBE). In contrast to density functional theory, which provides static and equilibrium properties, and molecular dynamics, known for its high computational demands and constraints related to timescale and system size, our NMC simulation of the PBE facilitates the direct observation of transient phonon transport within realistic device geometries subjected to Joule heating conditions. This approach like the phenomological non-local dual phase lag model \cite{Fotovvat2022,Roya2024,Sharif2025}, captures the spatiotemporal development of heat dissipation and hotspot formation in 2D materials, therefore allowing for a more practical assessment of their applicability in real-world thermal management and thermoelectric applications. Taking into account spatiotemporal 

At the first step in NMC simulation of the PBE, the phonons are mapped and distributed. Particularly, the phonons are allowed to move freely with their random velocity and direction, which are corresponding to the wave vectors in the phonon dispersion relations \cite{Mazumder2001,Shomali2017,2Shomali2017}. The studied materials have three acoustic and three optical branches. Importantly, the occupation of optical phonon states remains low at temperatures up to 600 K. Therefore, as confirmed by numerous studies, the contribution of optical branches to heat conduction is minimal and can be considered negligible, with only acoustic branches needing to be taken into account \cite{SMei2014}. Phonons in the optical branches, which have lower velocities and higher energies, are responsible for the formation of hotter regions. Therefore, in In-based monochalcogenides that reach temperatures above 600 K, such as InSe and In$_2$SeTe, neglecting these phonons could lead to a slight underestimation of the peak temperature. However, this is not a major concern in our analysis, as the focus is on identifying and categorizing the most efficient materials for thermoelectric and MOSFET applications. Therefore, in the present work, only the acoustic phonon branches are considered. Furheremore, the studied channels van taken to be almost thermally isotropic. Consequently, the contemplated acoustic phonon dispersion curves will be treated as the quadratic polynomials in the form of, $\omega_b$=c$_{b}$k$^{2}$+v$_ {b}$k \cite{Shomali2018}. The fitting coefficients for the quadratic equations are listed in the table. \ref{Tab1-tab1}.
	 
	 \begin{table*}[htbp]
		\caption{The fitting coefficients for four In-based materials of InSe, InSe under 4$\%$ and 6$\%$ strain, In$_2$SeTe and In$_2$SSe, which are embedded as a channel in a mosfet. The coefficient are obtained, discretizing the phonon dispersion curves reported in \cite{wang2019} and \cite{Vu2021}. }
		\label{Tab1-tab1}
		\vspace{-0.5cm}
\hskip -2cm
		 \centering
		\begin{center}
			\begin{small}
								\begin{tabular}{|p{2.2cm}|p{1.6cm}|p{1.6cm}|p{1.6cm}|p{0.6cm}|p{0.6cm}|p{0.6cm}|}
					\hline
					
	                Channel &  c$_{LA}$(m$^2$/s) & c$_{TA}$(m$^2$/s) & c$_{ZA}$(m$^2$/s) & v$_{LA}$ (m/s) & v$_{TA}$ (m/s) & v$_{ZA}$ (m/s)\\
					 \hline
InSe & {-4.061$\times$10$^{-8}$} & {-1.935$\times$10$^{-8}$}& {-5.443$\times$10$^{-9}$} & {484} & {278} & {179} \\
					 \hline
					4$\%$ Strained InSe &{-4.785$\times$10$^{-8}$} & {-2.063$\times$10$^{-8}$} &  {-1.101$\times$10$^{-8}$} & {572} & {335} & {245} \\ 
										 \hline
										 					6$\%$ Strained InSe &{-4.785$\times$10$^{-8}$} & {-2.063$\times$10$^{-8}$} &  {-1.101$\times$10$^{-8}$} & {572} & {335} & {245} \\ 
										 					\hline
										 In$_2$SeTe & {-4.061$\times$10$^{-8}$} & {-1.935$\times$10$^{-8}$} & {-5.443$\times$10$^{-9}$}& {484.2} & {278.6} & {179.2} \\
										 \hline
		In$_2$SSe & {-4.54$\times$10$^{-8}$} & {-1.92$\times$10$^{-8}$} & {-6.296$\times$10$^{-9}$}& {585.7} & {319.1} & {207.3} \\
					 \hline
				\end{tabular}
			\end{small}
		\end{center}
			\end{table*}
	
The interpretation of phonon heat transfer, in which the phonons act as the only heat transfer carriers, using the Boltzmann phonon transport equation (BTE) gives a better understanding of the thermal conductivity in our studied materials. The phonon BTE is a fundamental equation that can be used to describe the transfer of phonons in the nanoscale. This equation is solved to show the phonon distribution function evaluation over the time and space by considering the different phonon scattering mechanisms. The phonon BTE is expressed as,
         
 \begin{equation}
 \label{eq1}
 \frac{\partial f(\mathbf{r},\mathbf{k},t)}{\partial t} + \mathbf{v_g}(\mathbf{k}) \cdot \nabla f(\mathbf{r},\mathbf{k},t)= \left( \frac{\partial f}{\partial t} \right)_{\text{scatt}}.
 \end{equation}

In Eq. \ref{eq1}, the parameter $f(\mathbf{k},t)$, is the phonon distribution function that depends on the phonon wave vector, $\mathbf{k}$, and time, t.
Also, $\mathbf{v_g}(\mathbf{k}$) is the group velocity of the phonons and indicates how fast are the phonons with wave vector $\mathbf{k}$. The right side of the equation expresses the distribution function change caused by different phonon interaction processes, called scatterings \cite{Ge2016}. 

Here, the Umklapp phonon-phonon scattering rate is obtained from $\tau^{-1}_{b,U}(\omega)=\frac{\hbar \gamma ^{2}_{b}}{\bar{M} \Theta_{b} v^{2}_{s,b}} \omega^{2} T e^{-\Theta_b/3T}$, which is the standard general approximation for dielectric crystals \cite{THLiu2015}. In the mentioned formula, the parameter v$_{s,b}$ is the sound velocity of branch $b$ and $\bar{M}$ is the average atomic mass. Also, $\Theta$ is the Debye temperature. The exponential term presents the effective contribution from the redistribution caused by the N processes, while the remaining part indicates the standard Umklapp interaction strength. The values of $\Theta$, $\bar{M}$ and also, the Gr\"{u}neissen parameters, $\gamma_{TA,LA,ZA}$ are given in the table \ref{Tab2-BP}.  

	\begin{table*}[htbp]
		   \captionsetup{justification=centering}
	\caption{The parameters involved in phonon-phonon scatterings for the monolayers InSe, 4$\%$, In$_2$SeTe and In$_2$SSe taken from \cite{Wan2019} and \cite{Shafique2020}.  \newline}
	\label{Tab2-BP}
\vspace{-0.5cm}
\hskip -2cm
\centering
	\begin{small}

		\hspace*{-0.4cm}
		\begin{tabular}{|c|c|c|c|c|c|}
			\hline
			$Material$  & InSe & 4$\%$ Strained InSe  & 6$\%$ Strained InSe & In$_2$SeTe & In$_2$SSe \\
			\hline
			$\Theta$  & {111.96} & {90.0} & {90.0} & {98.0}  & {116.0} \\
			\hline
			$\bar{M}$ (e$^{-27}$ kg) & {160.837}  & {160.837.29}& {160.837}& {172.49} & {163.497} \\
			\hline
			$\gamma$ &  {0.51} & {0.64}& {0.64}& {0.99} & {0.7} \\
			\hline
		\end{tabular}
	\end{small}
\end{table*}

	\section{Numerical considerations}
	\label{Sec.4}

 At first, the whole frequency space for each acoustic branch is discretized. The maximum frequency in each polarization, $\omega_{max}$, is determined for each dispersion curve. The frequency interval between 0 and $\omega_{max}$ is divided into the N$_{int}$ intervals with $\omega_{o,i}$ and $\Delta \omega_{i}$ being, subsequently, the central frequency and the bandwidth of the $i-$th spectral bin. The phonon relaxation time, being obtained by inverting the scattering rates, are calculated with all 1000 frequency intervals in each branch. 
 
 Simultaneously, the velocity of all phonons within the 1000 different frequencies are obtained. Also, a uniform mesh of size 100 by 100 in the XY plane is considered. Then, by dividing the mesh size by the speed of each step, the phonon traveling time is found for each frequency interval. Finally, the minimum value of the calculated phonon traveling and relaxation times, is chosen as the time step. In accordance with the time step calculations, 3.399$\times$10$^{-12}$, 3.83$\times$10$^{-12}$, 3.78$\times$10$^{-12}$, 4.13$\times$10$^{-12}$ and 3.41$\times$10$^{-12}$ are, respectively, obtained for InSe, InSe under 4$\%$ and 6$\%$ strain, In$_2$SeTe and In$_2$SSe monolayers. Furthermore, the heat generation zone of Q=10$^{12}$ W/m$^3$, is contemplated at the center of the simulation area for a volume of 10$\times$100$\times$0.50 nm$^{3}$.  
	    
	\section{Results and Discussions}
	\label{Sec.5}
	
   \begin{figure}
  	\vspace{0cm}
  	\centering
  	\includegraphics[width=0.8\columnwidth]{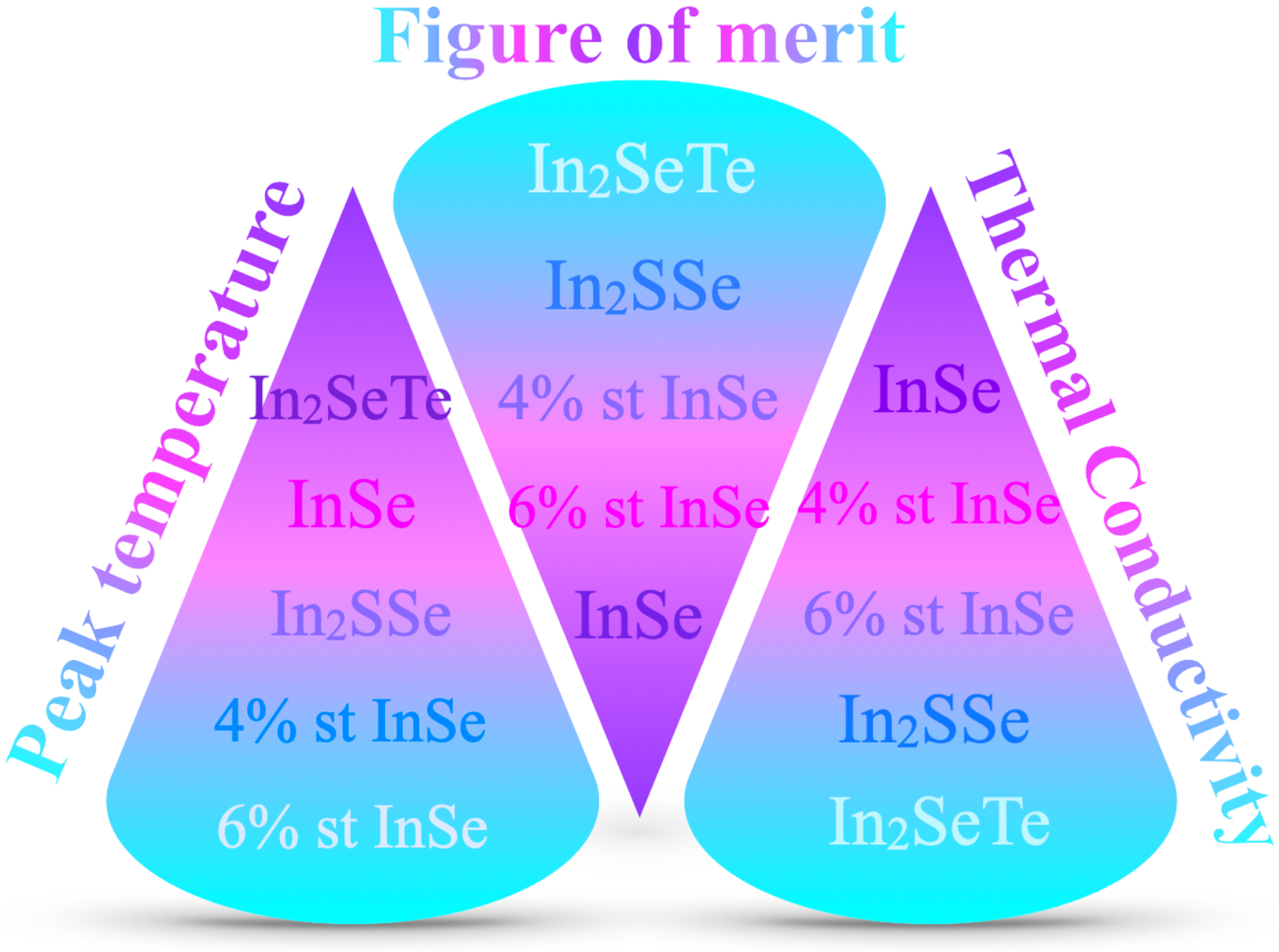}
  	\caption{\label{abstract} The maximum temperature, Figure of merit, zT, and thermal conductivity of different In-based monochalcogenides with 4$\%$ and 6$\%$ strained InSe and In$_2$SeTe being suggested, respectively, as the most thermally efficient channel for MOSFETs and thermoelectrically qualified material for TEGs.}
  \end{figure}
  
The transient heat transfer in the most famous In-based monolayers, is investigated using the non-equilibrium Monte Carlo simulation of the Boltzmann phonon transfer equation. The simulation has been run for 400 picoseconds. For the first 200 ps, the heat generation source is turned on, whereas, in the next 200 ps, the system is left to cool down. As presented in the Fig. \ref{abstract}, it is obtained that among the In-based family, the InSe under 4$\%$ and 6$\%$ strain presents the lowest maximum temperature. Also, In$_2$SSe, InSe, and In$_2$SeTe monochalcogenides, respectively, come with the next minimum peak temperatures. This is while the thermal conductivity is the least for In$_2$SeT and the greatest for InSe. This finding shows that the value of maximum temperature that the hotspot in a monolayer channel obtains, not only depends on the behavior of the frequency versus the wave vector, but also is very sensitive to the phonon-phonon scattering rates. 

Reaching the lowest maximum temperature while being under the influence of heat generation zone, means that the transistor with channel of InSe with $4\%$ strain, is the most reliable and thermally efficient one. Simultaneously, as also seen in the Fig. \ref{abstract}, the Janus monolayer of In$_2$SeTe with the maximum peak temperature and the minimum lattice thermal conductivity has the highest zT figure of merit which makes it the most suitable material among the studied In-based monochalcogenides, for being used as the thermoelectric material in TEGs.
    
    \begin{figure}
 	\vspace{0cm}
 	\centering
 	\includegraphics[width=\columnwidth]{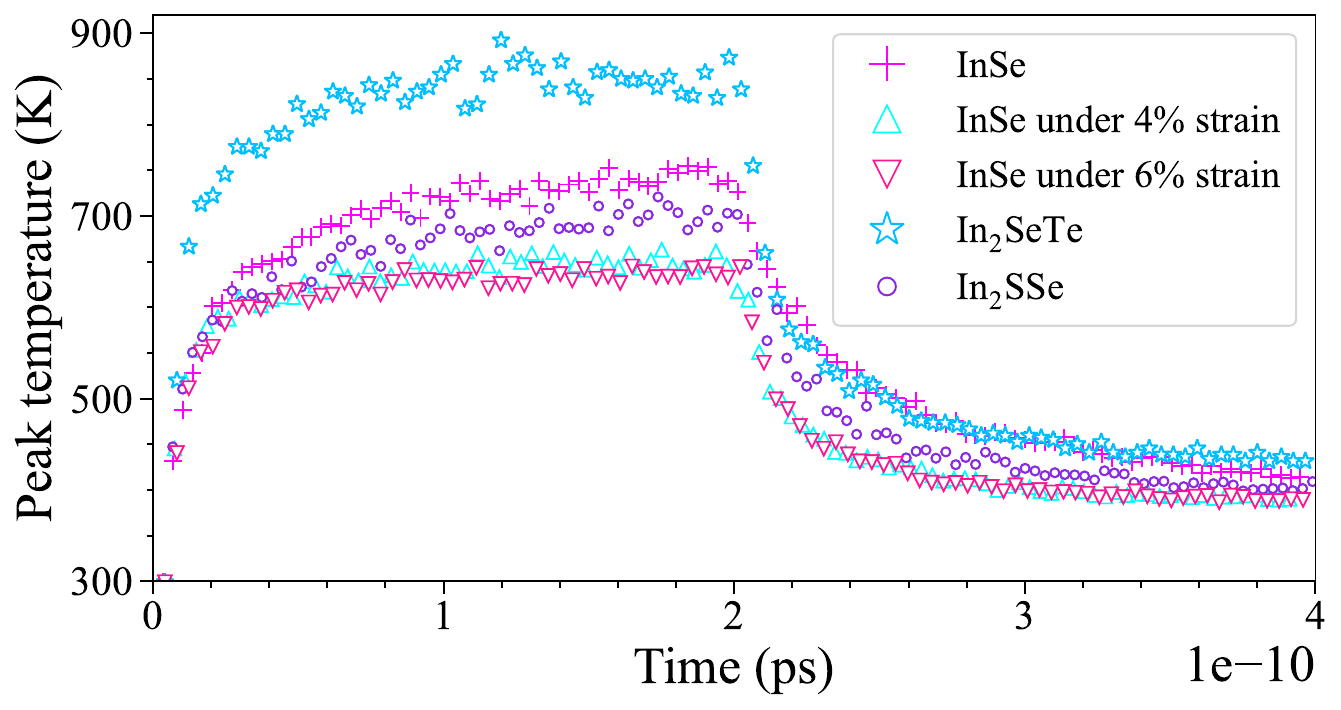}
 	\caption{\label{peaktemp} The maximum temperature versus time at the XY plane experienced within the 400 ps simulation. The markers +, $\triangle$, $\triangledown$, $\star$ and $\circ$ markers, respectively, demonstrate the result for InSe, InSe under 4$\%$ and 6$\%$ strain, Janus In$_2$SeTe and In$_2$SSe. The InSe under 6$\%$ strain and Janus In$_2$SeTe 2D material has reached, subsequently, the lowest and the highest peak temperature and can be proposed, respectively, as the very thermally efficient candidate for silicon channel replacement and the thermoelectric material for TEGs.}
 \end{figure}
 
 In the following, more details about the thermal transport through the In-based monolayers will be given. As the reliability of a transistor is determined by the value of its maximum temperature, firstly, the peak temperature rise demeanour is precisely scrutinized. The state of maximum temperature for all five In-based materials of InSe, InSe under 4$\%$ and 6$\%$ strain, Janus In$_2$SeTe and In$_2$SSe, is presented in Fig. \ref{peaktemp}. Among the studied materials, two Janus monolayers of In$_2$SeTe and In$_2$SSe were expected to present the highest maximum temperature as these materials are thermomechanically favorable monolayers. This is due to their more complicated unit cells which make much more phonon scatterings \cite{Bahadori2024}. As is evident in Fig. \ref{peaktemp}, the maximum temperature that Janus monolayer In$_2$SeTe, reaches, is significantly higher than the other studied InSe monochalcogenides, which is alongside our expectation. Meanwhile, in contrast to what we anticipate, In$_2$SSe exhibits a thermal profile between that of the 2D strained and normal InSe monolayer. 
 
 Although, this finding presents that the In$_2$SSe monolayer with an acceptable bandgap (1.1–1.5 eV), and a good carrier mobility (885 cm$^2$/V/s), can be an efficient candidate for using as a channel in MOSFETs rather than being utilized in TEGs as a thermoelectric material; the high Seebeck coefficient alongside the almost low thermal conductivity of In$_2$SS, makes this material also a strong candidate for thermoelectric engineering. It should be mentioned that nevertheless the lattice thermal conductivity of In$_2$SS is low, but the total thermal conductivity, considering also the low optical phonons, is comparable to that of the InSe monolayer as the dispersion curve behavior for the two 2D layers is very similar. In more detail, the $\kappa$ of monolayer InSe and In$_2$SSe is similar to the semiconductors GaAs, few-layer MoS$_2$, and few-layer black phosphorus. In contrast, the monolayer In$_2$SeTe with heavier atomic mass has a much lower $\kappa$ \cite{Wan2019}. This is why the In$_2$SS monolayer presents a much lower maximum peak temperature relative to the 2D In$_2$SeTe material and is more practical for thermal management solutions.
 
 \begin{figure}
	\hspace{-1.0cm}
	\centering
	\includegraphics[width=1.05\columnwidth]{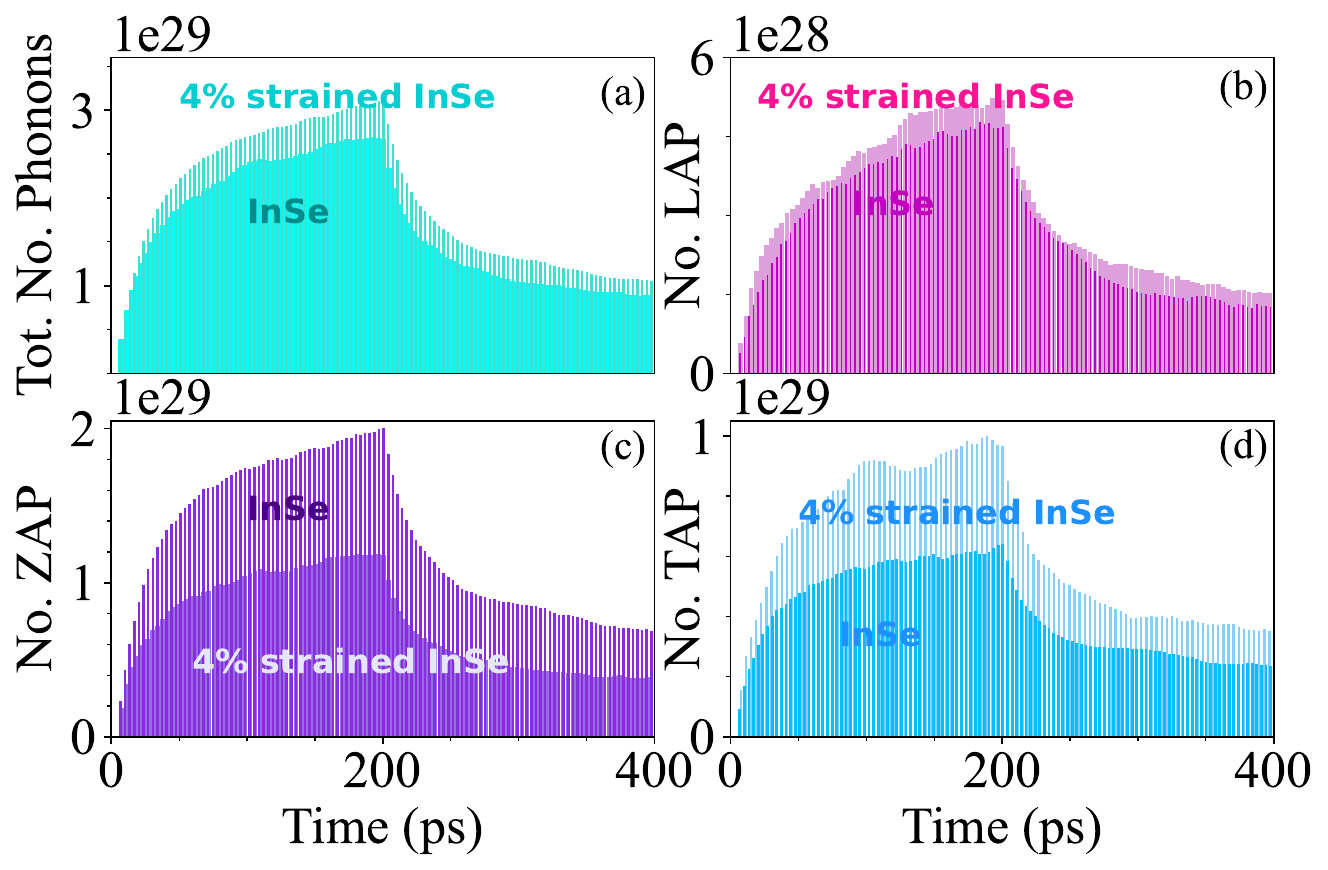}
	\caption{\label{phononanalysisst-unst} The time-dependent behavior of the (a) total number of phonons, the number of (b) LA, (c) TA, and (d) ZA phonons, for two unstrained/4$\%$ strained InSe monolayer materials, presenting while the LA phonons for monolayers are almost the same, the number of TA and ZA phonon are, respectively, doubled and halved for strained InSe. This confirms the lower maximum temperature for strained monolayers. }
\end{figure}

\begin{figure*}
	\vspace{0cm}
	\centering
	\includegraphics[width=1.6\columnwidth]{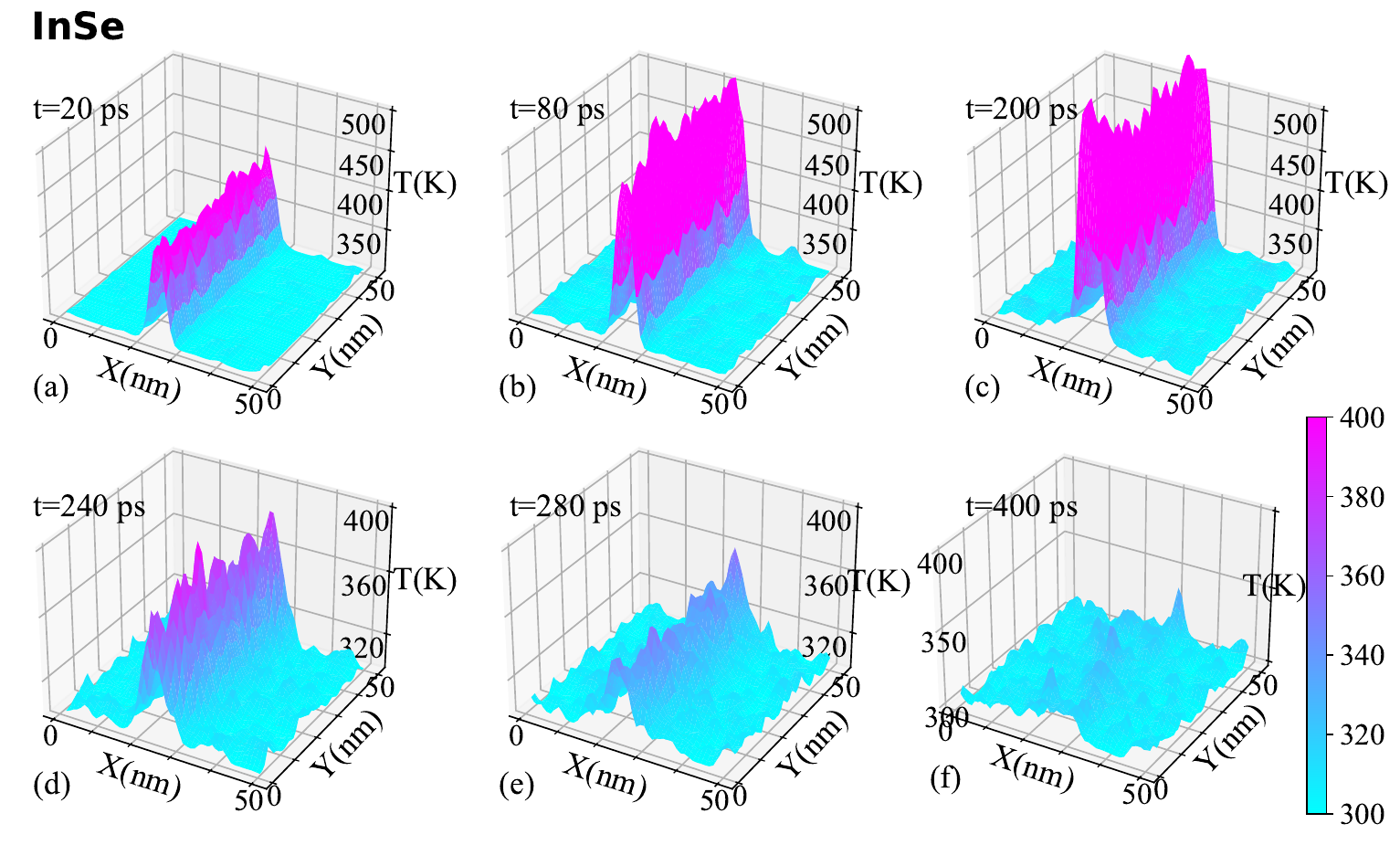}
	\caption{\label{profINSE} The hotspot temperature behavior on XY plane during the heating and cooling phases when t= (a) 20, (b) 80, (c) 200, (d) 240, (e) 280, and 400 ps, for InSe monolayer. The average temperature of the hotspot, appearing instantly, increases upto 600 K at 200 ps. While the heat generation zone is turned off, the cooling trend of monolayer is very slow, such that at time t=400 ps, it is still hot. Both behavior are attributed the aggregation of the ZA phonons.}
\end{figure*}

 \begin{figure*}
	\vspace{-1.2cm}
	\centering
	\includegraphics[width=1.6\columnwidth]{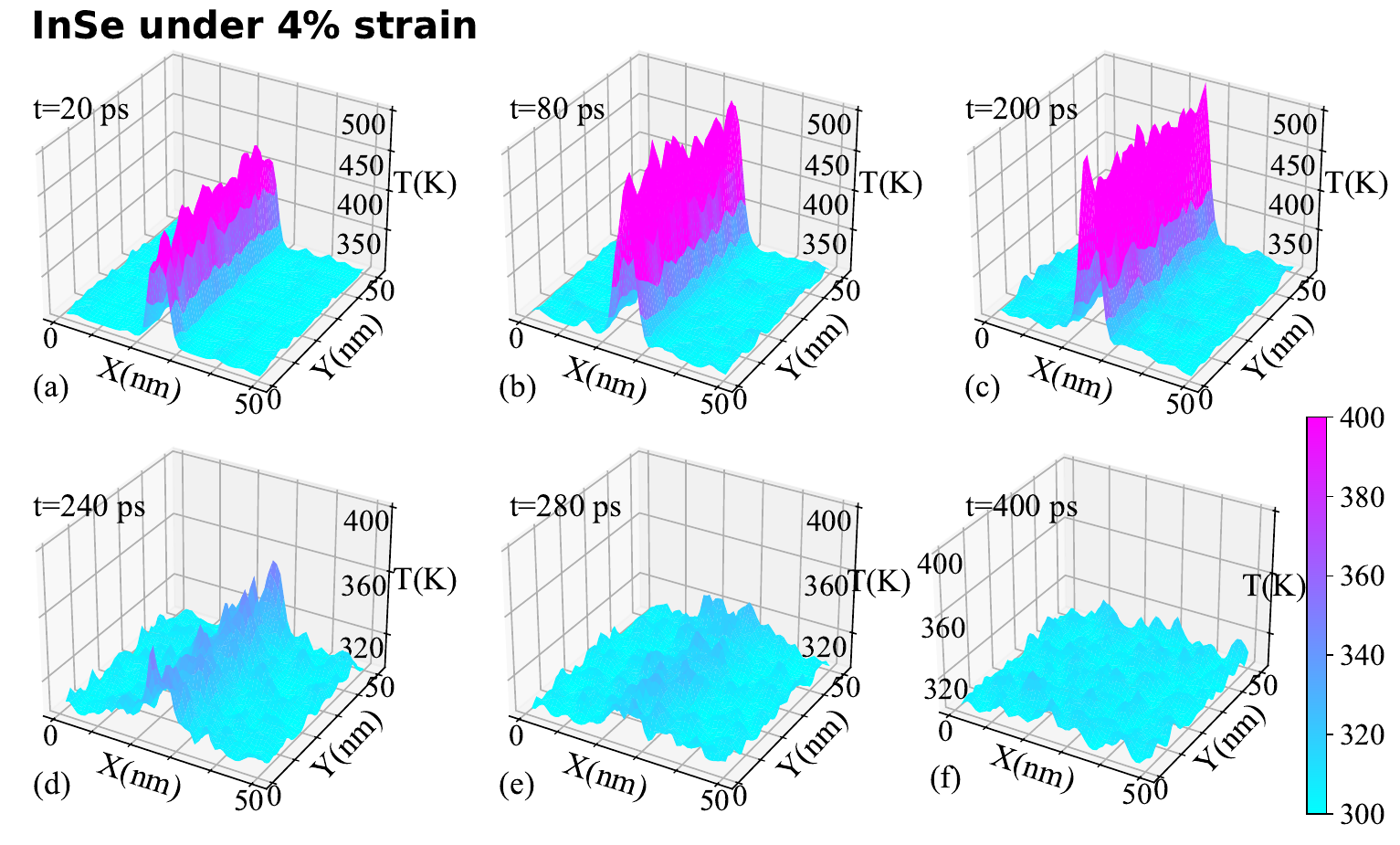}
	\caption{\label{profINSEstrain4} The In-plane temperature profile during the heating to cooling phases when t= (a) 20, (b) 80, (c) 200, (d) 240, (e) 280, and 400 ps, for the 2D material of InSe under 4$\%$ strain. The plots present the maximum average temperature of hotspot is at least 100 K less than that of pristine InSe. Also while at t=240 ps, the hotspot in pristine InSe is still vey hot having the average temperature of 400 K, the strained InSe is reaching the steady state.}
\end{figure*}

 \subsection{Strain Effects}
 It merits attention that the behavior of hotspots for each In-based dichalcogenides is different as the contribution of longitudinal acoustic, LA, transverse acoustic, TA, and out-of-plane acoustic, ZA, phonons differs. As previously mentioned, the peak temperature is the lowest for the material, strained InSe. It is worthwhile to look at the phonon analysis to uncover the physics behind. Figure \ref{phononanalysisst-unst} presents the behavior of the number of phonons over time for both unstrained and strained InSe monolayers. While the total number of phonons for unstrained and strained InSe two dimensional materials are nearly the same, the number of the ZA and TA phonons for the two monolayers differ notably. Also the number of the LA phonons for InSe and 4$\%$ strained InSe varies to some degree. As seen, the contribution of the ZA and TA in both In-based dichalcogenides is dominant. In more detail, while the heating region is on, during the first 200 picoseconds, the number of slow ZA phonons for strained InSe is approximately half that of InSe. Meanwhile, the number of almost fast TA phonons, say 0.5$\times$10$^{29}$ phonons in unstrained InSe, is half that of 4$\%$ strained InSe. Furthermore, the number of fast LA phonons, which have the least contribution to heat transport in this context, is approximately the same for both unstrained and strained InSe monolayers. The ZA phonons are slow, so when they are trapped in the hotspot region, they hardly move away from it. Therefore, more ZA phonons leads to increased trapped phonons and consequently higher temperature. The TA phonons, which have lower frequencies than the LA phonons and relatively higher velocities compared to the ZA phonons, possess lower energy but are sufficiently fast to escape the hotspot region. Consequently, more TA phonons can be associated with the reduction in the peak temperature. Conversely, the LA phonons exhibiting higher frequencies and velocities, with these parameters working in an opposing manner in determination of a peak temperature, are almost the same for both monolayers. According to this explanation, while the number of the LA phonons are low and almost the same for both strained/unstrained InSe, the simultaneous presence of twice the number of the TA phonons and half the number of the ZA phonons in strained InSe monolayer, is responsible for lower peak temperature achieved during the heating in this two-dimensional material.
 
 Further, the more detailed temperature profile behaviors of strained/unstrained InSe are presented in Figs. \ref{profINSE} and \ref{profINSEstrain4}. As seen, the hot spots are formed at the heat generation zone where the hot phonons are injected into the transistor. As discussed previously, owing to the existence of more energetic and less fast phonons in an InSe monolayer, the temperature all over the 2D material is always 100-200 K higher relative to that of the InSe under 4$\%$ strain. The same trend is also seen during the cooling process of the InSe monolayers. With 2$\times$10$^{29}$ ZA phonon per volume being the dominant contribution of the phonons, it takes much more time for pristine InSe to settle down in a steady state while the Indium selenide under strain cools down earlier. For instance, the temperature profile for a strained InSe at T=240 ps is still lower than that of the pristine InSe even at t=280 ps. Finally, at t=400 ps, while the InSe experiences temperature as hot as 350 K at some points, the strained InSe have reached the steady state for almost 50 ps ago, as a consequence of having more TA, and less ZA phonons. Indium Selenide under strain cools down earlier due to the fact that it has a lower temperature during the heating time and is less heated.
 
 \begin{figure}
 	\centering
 	\includegraphics[width=1.05\columnwidth]{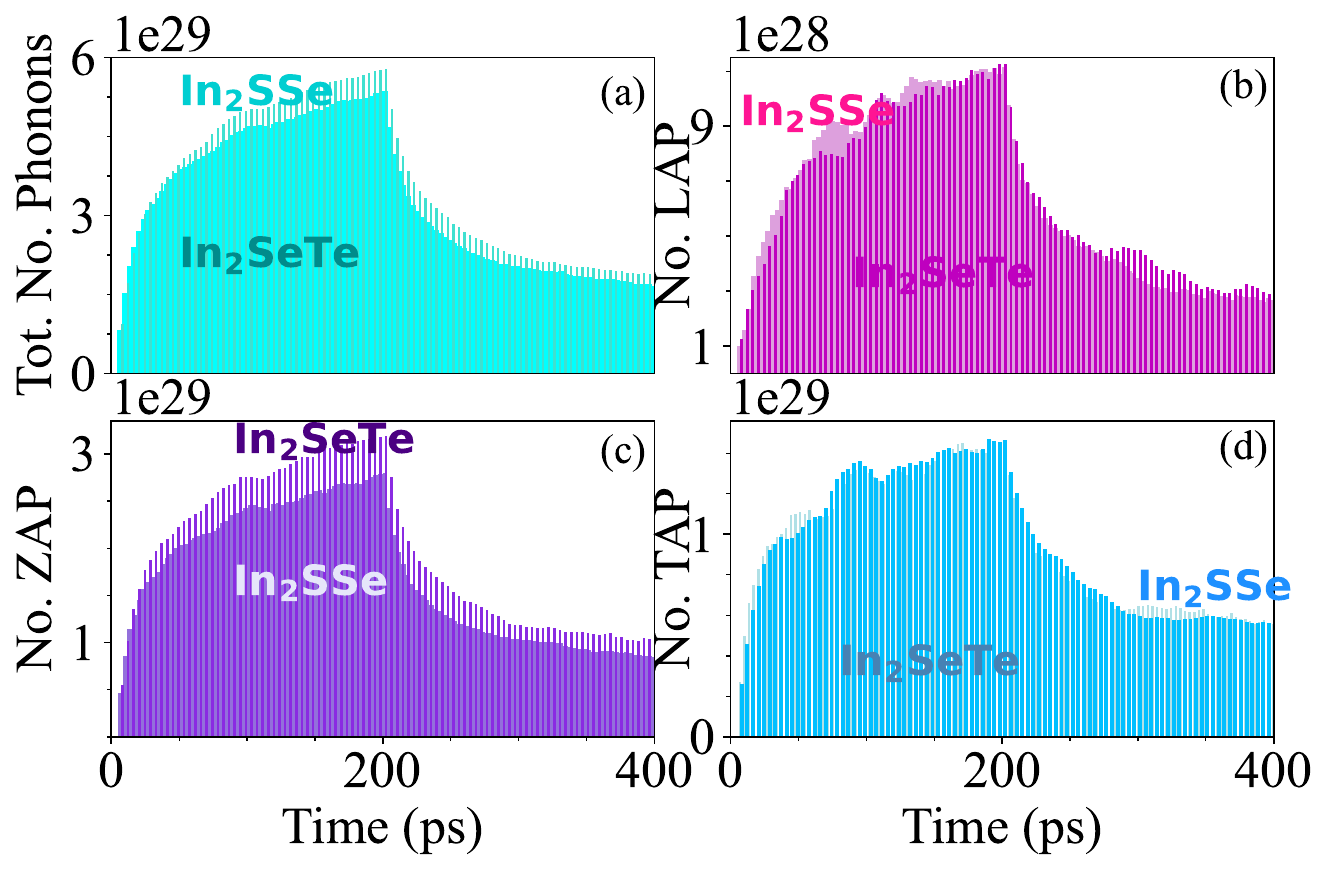}
 	\caption{\label{la-ta-za-phonons-2} The Demeanor of the (a) total number of phonons, the number of (b) LA, (c) TA, and (d) ZA phonons, versus time, for two Janus In$_2$SeTe and In$_2$SSe monolayers, confirming the number of ZA phonons for In$_2$SeTe is consistently 16$\%$ higher than that of the In$_2$SSe.}
 \end{figure}

 \begin{figure*}
 	\vspace{0cm}
 	\centering
 	\includegraphics[width=1.6\columnwidth]{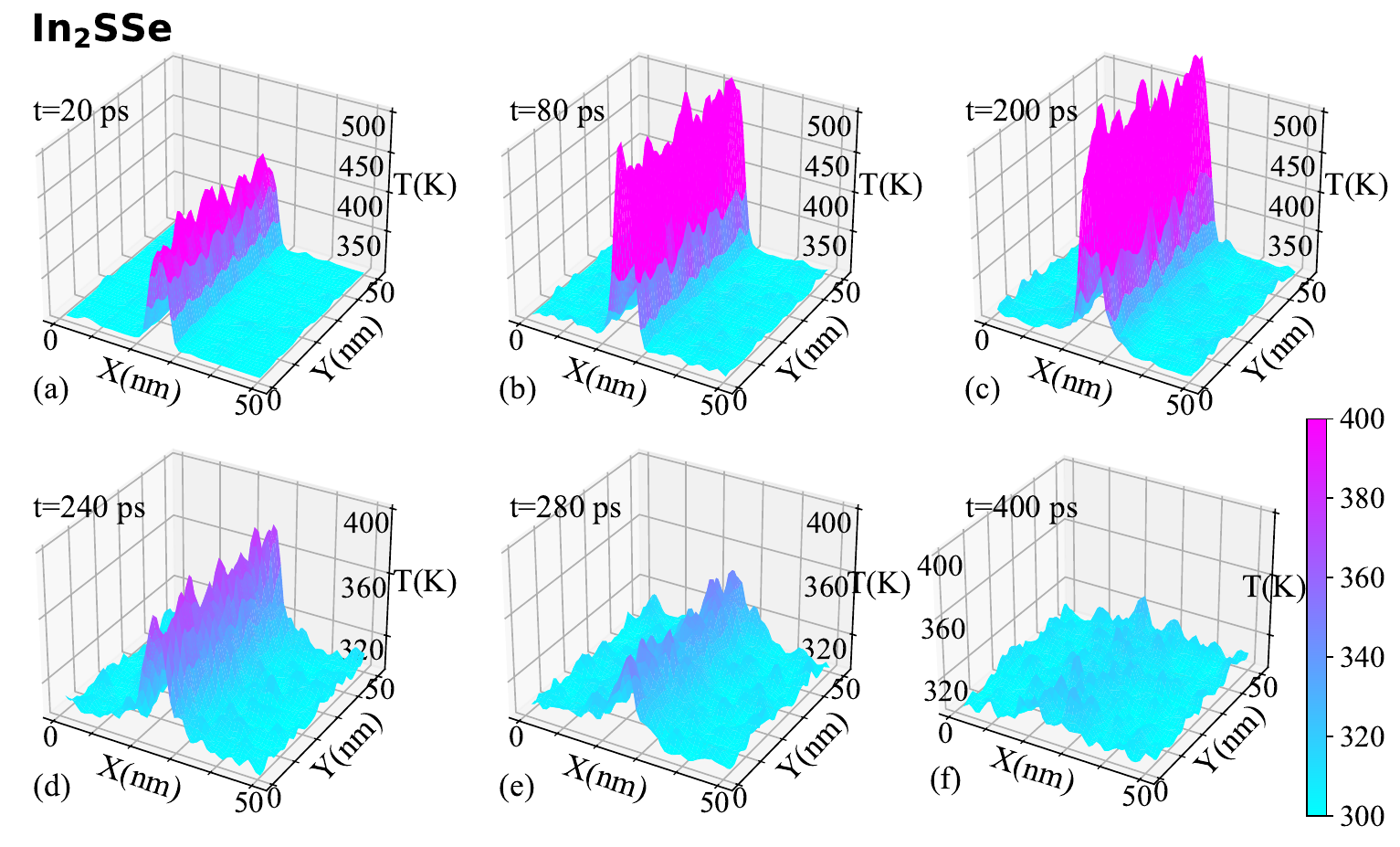}
 	\caption{\label{profIN2SSE} The temperature distribution for 2-D material of In$_2$SSe at  t= (a) 20, (b) 80, (c) 200, (d) 240, (e) 280, and 400 ps.}
 \end{figure*}
 
 \subsection{In-based Janus Monolayers}
\begin{figure*}
	\vspace{0cm}
	\centering
	\includegraphics[width=1.6\columnwidth]{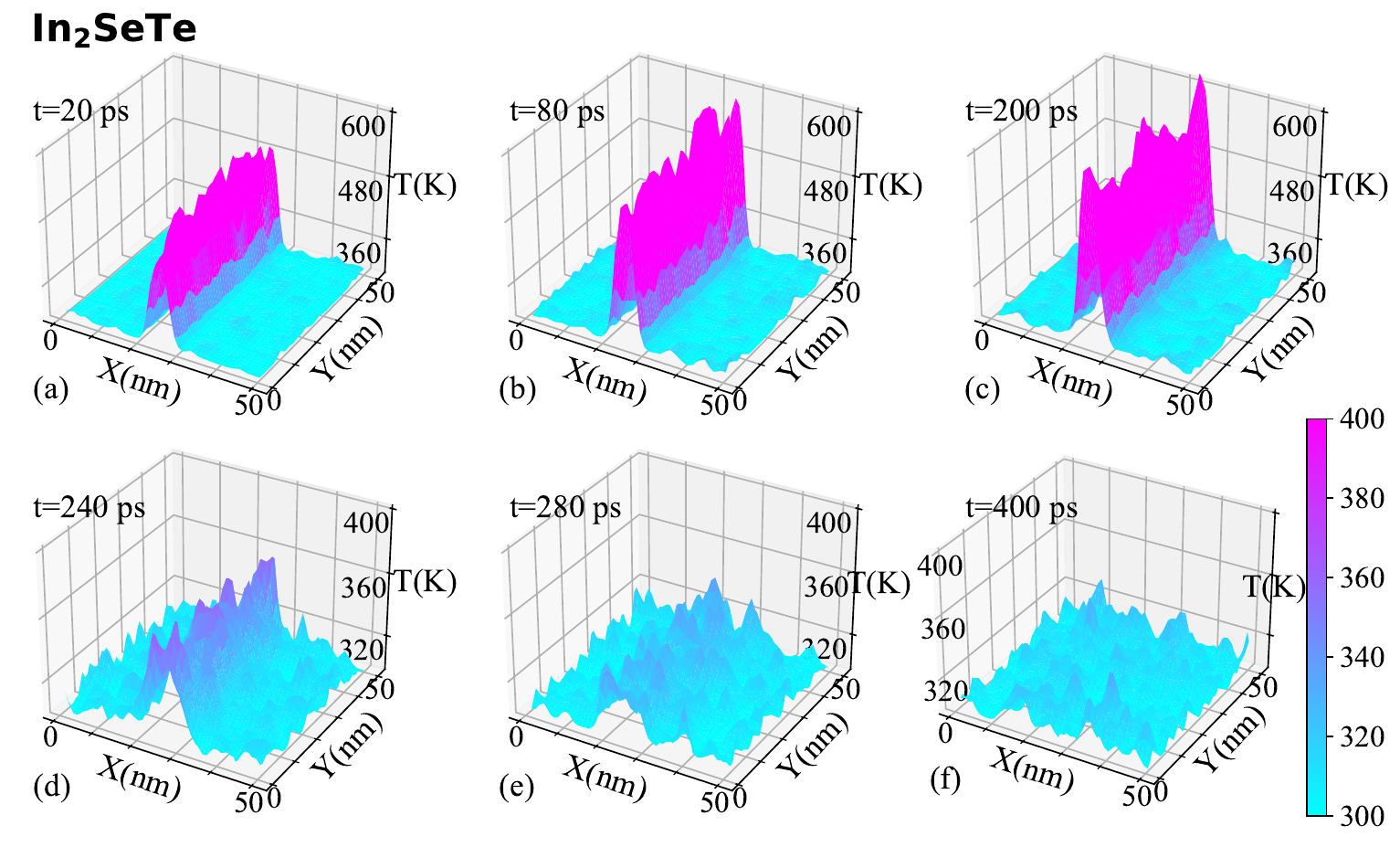}
	\caption{\label{profIN2SETE} The same as \ref{profIN2SSE} but for the In$_2$SeTe monolayer, presenting consistently the higher maximum temperature relative to the janus In$_2$SSe material.}
\end{figure*}

At the next step, the phonon analysis for the studied Janus In$_2$SeTe and In$_2$SSe monolayers is given. Figure \ref{la-ta-za-phonons-2} demonstrates the number of phonons in each branch versus time. The phonons contributed to the heat transport are mostly the phonons in ZA and TA branches. In addition, the Janus In$_2$SeTe and In$_2$SSe have nearly the same number of the TA and LA phonons. The ZA phonons are much slower than the LA and TA phonons moving through the transistor. The slow phonons do not take the heat from the hot region to the surroundings and consequently the heat falls into the trap and the hotspots are formed. Accordingly, as already discussed, the Janus In$_2$SSe with less ZA phonons is expected to reach lower maximum temperature. This expectation is confirmed by our calculations, as shown in Figs. \ref{profIN2SSE} and \ref{profIN2SETE}. These figures illustrate the temperature distribution behavior at the XY plane for two monolayer Janus structures of In$_2$SeTe and In$_2$SSe. As depicted, given that the monolayers are subjected to the Joule heating mechanism over a period of 200 ps, the immediate formation of hotspots are clearly recognizable. Figures \ref{profIN2SSE} and \ref{profIN2SETE} (a-c) present that during the heating, the temperature profile in overall structure is, on average, 100 K lower than the In$_2$SeTe temperature. Also during the 200 ps of cooling, the same tendency is also noticeable. This difference is such that at t=400 ps, while the In$_2$SeTe monolayer still exhibits temperatures 50-100 K higher than the ambient, almost the entire In$_2$SSe structure has reached thermal relaxation.

\begin{figure}
 	\vspace{0cm}
 	\centering
 	\includegraphics[width=0.8\columnwidth]{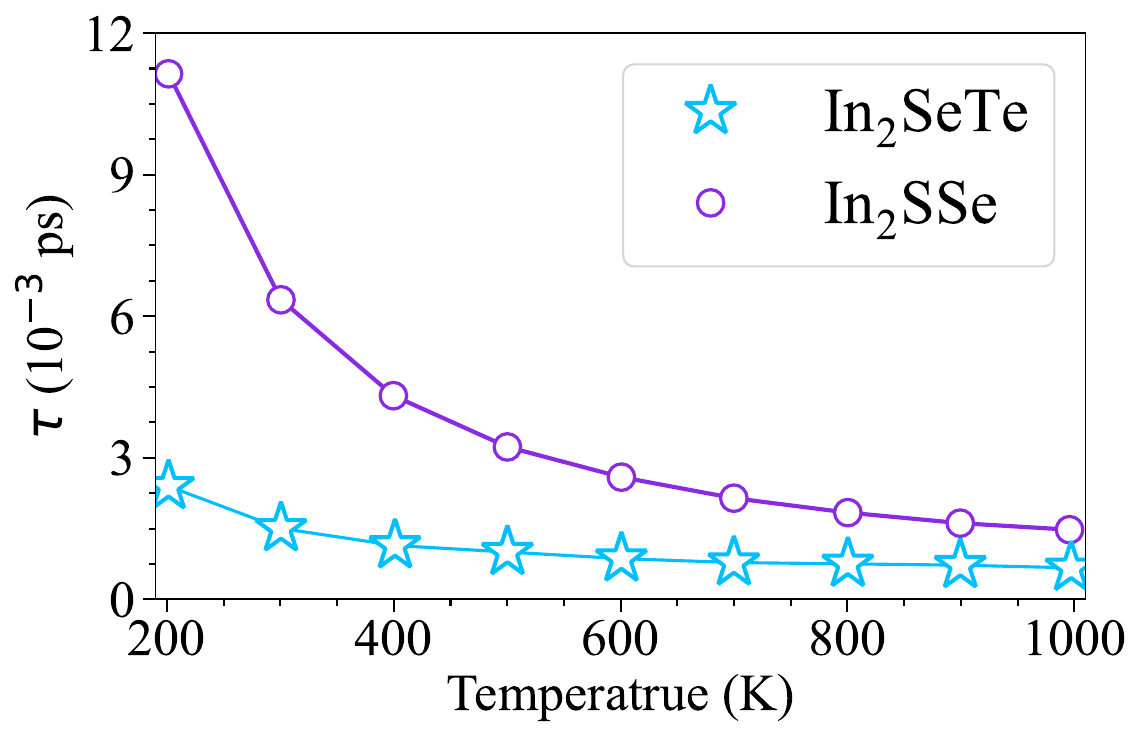}
 	\caption{\label{scatteringJanus} The temperature dependent relaxation time $\tau$ of the Janus In$_2$SSe and In$_2$SeTe monolayers (the plots are reproduced from the work \cite{Vu2021}.)}
 \end{figure}

Alongside the phonon distribution arrangement in advantage of lower peak temperature for In$_2$SSe, the significant temperature difference can also be attributed to the phonon-phonon interactions. In more detail, the temperature dependent relaxation time $\tau$, for the two studied Janus monolayers, is demonstrated in Fig. \ref{scatteringJanus}. As shown, the relaxation time for In$_2$SSe is almost four times that of the In$_2$SeTe. Hence, the scattering rate, which represents the strength of the interactions of the phonons and in RT approximation is calculated as $1/\tau$, is much higher for Janus In$_2$SeTe \cite{Vu2021}. A higher scattering rate restricts the ability of phonons to efficiently transport heat and so reduces the thermal conductivity of the material. The thermal conductivity decrease caused by the increased phonon scattering, prevents the heat to be transported away from the hot zones. Consequently, the localized heating engendered from the external sources accumulates instead of dissipating. The accumulation of energy increases the internal energy of the 2D layer and this is also the reason why the temperature of the In$_2$SeTe rises more.

  \subsection{Thermoelectric material/transistor channel selection}
To put it concisely, among the five studied In-based monolayers, strained InSe exhibits the lowest peak temperature under self-heating conditions, while the Janus In$_2$SeTe reaches the highest. Since transistor reliability is primarily governed by the maximum temperature at thermal hotspots, rather than the average die temperature, strained InSe emerges as the most suitable candidate for replacing silicon channels in MOSFETs. Conversely, the high peak temperature and resulting large temperature gradient observed in In$_2$SeTe make it a promising material for thermoelectric generator applications. 
 

	\section{Conclusions}
        \label{Sec.6}

Heat transport in five indium-based monolayers was investigated using nonequilibrium Monte Carlo simulations of the phonon Boltzmann transport equation (PBTE). The PBTE framework, which incorporates phonon distribution functions, group velocities, and scattering mechanisms, enables accurate prediction of nanoscale thermal behavior in low-dimensional materials. Our results indicate that the InSe monolayer possesses relatively high thermal conductivity, making it a promising candidate for use as a transistor channel material. In contrast, the Janus In$_2$SeTe monolayer exhibits lower thermal conductivity but enhanced thermoelectric properties, rendering it more suitable for energy harvesting applications. Notably, the study highlights the significant impact of strain engineering in modulating phonon transport in In-based dichalcogenides, offering an effective strategy for optimizing thermal performance in nanoelectronic devices. Specifically, we find that MOSFETs incorporating strained InSe channels exhibit reduced peak temperatures and faster cooling rates compared to unstrained InSe, confirming its thermal superiority for transistor applications. On the other hand, the Janus In$_2$SeTe monolayer, characterized by higher peak temperatures and strong temperature gradients, emerges as an excellent thermoelectric candidate for use in thermoelectric generators. In conclusion, our findings support the strategic use of strain-engineered and Janus-type monolayers in the design of next-generation electronic and semiconductor devices, enabling improved thermal management and energy efficiency.

\end{document}